\documentstyle[12pt]{article}
\begin{document}

\baselineskip=24pt
\setcounter{page}{1}
\parskip=0pt plus2pt
\textheight=22cm

%%%%%%%%%%%%%%%%%%%%%%%%%%%%%%%%%%%%%%%%%%%%%%%%%%%%%%%%%%%%%%%%%%%%%%%%%
\begin{titlepage}

\begin{center}
{\Large\bf 
Temperature dependence of the Power law exponent of 
relaxation in a supercooled Liquid }
\end{center}

\vspace{0.2in}

\begin{center}
{\it Sudha Srivastava and Shankar P. Das 
\\ School of Physical Sciences, Jawaharlal Nehru
University
\\New Delhi 110067, India.}

\end{center}
\vspace*{1.0in}

\begin{center}
{\bf ABSTRACT}
\end{center}
\noindent
The dynamics of Lennard-Jones fluid is studied through
extended mode 
coupling theory (MCT) with the inclusion of the slow
mode of defect
density. Inclusion of defect density facilitates the
liquid like state for 
temperatures much lower than predicted from ideal MCT.
From the present model
 the temperature dependence of the power law exponent
is obtained at 
a {\em constant pressure}. We have also computed
the wave number dependence of the power law exponent.

\noindent

\vspace{1in}
\noindent
PACS number(s) : 64.70P, 05.60, 64.60C  
\end{titlepage}

\newpage

\section*{Introduction}
The phenomena of Glass transition has been studied
extensively through theoretical 
\cite{freevol,forster,kimmaj,sid-trans,leshouc}
models, computer simulations 
\cite{ullo-yip,hiwa} as well as scattering experiments
\cite{megen,van-uwood,van-pusey}.
 If a liquid is cooled beyond the freezing point
avoiding 
the first order transition to an ordered crystalline
 state, it continues in the amorphous state with solid
 like properties. In this supercooled state the
 liquid develops very long relaxation times
 and the transport coefficients (viscosity) abruptly
increase. 
 A microscopic mechanism for this slow dynamics using a statistical 
 mechanical approach has been proposed through the Mode coupling 
 theory (MCT). 
 This is based upon the idea \cite{beng} that the transport 
coefficients 
 in the supercooled state gets a feedback from the nonlinear coupling 
 of slow modes in the fluid.
The effect of the structure on the dynamics is studied from the wave
  vector dependence of the model.  MCT predicts a two step
relaxation process for glassy dynamics : power law
($\beta$-relaxation) at 
intermediate time scales which crosses  over to
stretched exponential 
($\alpha$-relaxation) decay  of density correlation
function over long time 
scales. In the ideal model  \cite{leu} there is a sharp 
transition from ergodic to a nonergodic phase at a
critical density.
In the later version of more careful analysis 
\cite{D+M,jim,gt-sj} the sharp
transition is finally 
removed from the ergodicity restoring mechanisms
coming from the 
coupling of the density fluctuations with currents in
the compressible
fluid. The final ergodicity restoring mechanisms in a
supercooled
liquid behaves like a collective slow mode. There has
been phenomenological extension of the theory to consider extra slow 
modes that develop in the liquid in the supercooled region. 
In the present work we consider an extension \cite{dsch,spd,yeomaj,yeo} 
of
the simple feed back mechanism of ideal MCT through the inclusion
of the extra slow mode of defect density in the hydrodynamic
description for the isotropic liquid.
Here we use this model to study the dynamics in the
simple One component Lennard-Jones fluid in the supercooled
state.
In this respect we like to note that previous MCT
studies on LJ liquid \cite{beng1} show that one component system
freezes into a nonergodic phase at a relatively low density.

The defect density considered in the model can be
associated with the 
 vacancies in the solids or void or free volumes in
the fluid. 
 The movement of these free volumes were considered
crucial for the 
transport in the liquid \cite{freevol}.
Following Ref. \cite{yeo} we consider the defects to
be moving in a metastable potential well. 
Here the extent of coupling between defect density and the
particle density governs the location of the ideal glass transition 
point that occurs as a result of the feedback mechanism proposed in 
MCT.
Thus at a constant pressure  we observe
a transition line whereas a transition point is
predicted in
 the simple MCT. Phase
diagram of coupling parameters with temperature is
studied through 
nonergodicity parameter. We compute the relaxation
dynamics of
fluid in the power law regime and calculate the power
law exponents. From
the present model the temperature dependence of the
relaxation exponents
arises naturally as a consequence of the critical
line.
The paper is organized as follows. In next section  we
briefly describe the
model considered followed by section III where we
discuss briefly
 the static strucutre factor used in the calculation. 
In section IV the results for the power law relaxation
as
predicted by the mode coupling theory is presented. 
In last section a brief discussion of the work is
presented.

\section*{The Model Studied}
In the mode coupling theory we focus on the 
normalized density-density correlation function
$(\psi_q(t))$.
The dynamical evolution of $\psi_q(t)$ is written in
terms of an 
integro-differential equation,
\begin{equation}
\label{dynamic}
\Delta_q \ddot{\psi}_q(t) + \Gamma_q^0 \dot{\psi}_q(t) + \psi_q(t)
+ \int_0^t ds \Gamma_q^{mc}(t-s) \dot{\psi}_q(s) =0,
\end{equation}
\noindent
where the constant $\Delta_q=\beta m S(q)/q^2$ with $m$ as the mass of the liquid particle 
and $\beta$ is the Boltzman factor. S(q) is the static structure factor for the liquid.
The bare transport coefficient $\Gamma_q^0$ \cite{kirk} is defined in a dimensionless form as 
\begin{equation}
\Gamma_q^0 = \frac{2 \Delta}{3t_E}\left[1-j_0(q^*)+2j_2(q^*)\right], 
\end{equation}
where $q^*=q\sigma$ and $j_n(x)$ is the $n^{th}$ order spherical Bessel
function \cite{abramowitz}. The Enskog time $t_E$ \cite{PY}
for the LJ fluid is approximated  using the equivalent hard sphere
diameter.
The generalized transport coefficient $\Gamma_q(t)$
has the
bare ($\Gamma_q^0$) as well as the mode coupling ($\Gamma_q^{mc}(t)$) 
contributions from the nonlinear coupling of density fluctuations.
Following Ref. \cite {yeo},
the mode coupling part of the generalized transport
coefficient  is given by,
\begin{equation}
\label{visd}
{\Gamma}_q^{mc}(t) = \int \frac{d{\vec k}}{(2\pi)^3}
[\tilde{V}^{(1)}(q,k)\psi_k(t) \phi_{k_1}(t) + \tilde{V}^{(2)}(q,k)\psi_k(t) \psi_{k_1}(t)]
\end{equation}
\noindent
with $\vec{k_1} = \vec{q} - \vec{k} $. Here $\phi_{k_1}(t)$ is the
 normalized defect auto correlation function and gives the
 contribution to the transport coefficient coming from the
 coupling of the defect density with the density fluctuations.
The mode coupling  vertices with full wave
vector dependence 
are  given by
\begin{equation}
\label{v1}
\tilde{V}^{(1)}(q,k)={\frac{S(q)}{n_0}} \big[ 2y  \tilde{U}(q,k)S(k)S(k_1)
+ \kappa n_0c(k)S(k) \big]
\end{equation}
\noindent
and
\begin{equation}
\label{v2}
\tilde{V}^{(2)}(q,k)=\frac{S(q)}{2 n_0} \big[\tilde{U}^2(q,k)
-2 y \tilde{U}(q,k) \big]S(k)S(k_1)
\end{equation}
\noindent
These equations are obtained \cite{yeo}
 from the nonlinear equations for the extended set of 
hydrodynamic variables.  The vertex function $\tilde{U}$ in
(\ref{v2})
 is given by,
\begin{equation}
\label{u}
\tilde{U}(q,k) = \frac{n_0}{q} \big[ \hat{q}.k c(k) +
\hat{q}.(k_1)c(k_1) \big]
\end{equation}
\noindent
where $n_0$ is the particle number density. 
It is assumed  that the defects 
are moving in a metastable double well potential with
$\kappa$ and $y$ 
as two dimensionless parameters characterizing the
coupling of the
 defect densities with the particle density and the
depth of the potential 
well. Defect density is assumed to be 
weakly interacting with the mass density and is
treated as a variable 
similar to mass density.
For $\kappa=y=0$, (\ref{visd}) reduces to the standard
result,
\cite{beng}.
The presence of the linear term in the mode coupling
 vertex comes through the coupling with the very
slowly varying defect 
correlation \cite{bongsoo}.
In the next section we study the implications of this
feed back mechanism
 that arises from the coupling of the density
fluctuations as well as
 the defect density coupling. For this the static
structure factor is 
 taken as an input - measuring the extent of
structural effects on
  relaxation. We use the standard model \cite{wca}
for the Lennard-Jones Structure factor. We give a brief description of 
computation of the static structure factor next.
\section*{Static Structure Factor}
In mode coupling equations interparticle interaction 
potential enters 
as an input via static structure factor $S(q)$. 
For a Lennard-Jones (LJ) fluid the 
perturbation approach described in \cite{anderson} is
being
 used for the computation of the structure factor. 
The full Lennard-Jones potential,

\begin{equation}
U(r)=4\epsilon[(\sigma/r)^{12} - (\sigma/r)^6]
\end{equation}
is divided into a repulsive  part $U_0(r)$ and an
attractive part $U_A(r)$ as,
\\
\[ U_0(r) = \left \{ \begin{array}{ll}
U(r) + \epsilon, &~~~~~~r< r_0 \\
0,  &~~~~~~r\ge r_0
\end{array}
\right. \]
\[U_A(r) =\left \{ \begin{array}{ll}
-\epsilon  &~~~~~~r< r_0\\
U(r), & ~~~~~~r\ge r_0
\end{array}
\right. \]
\\
with $r_0=2^{1/6}\sigma$. A trial system with hard
sphere diameter 
$d\le r_0 $ and an attractive potential energy part
$U_A(r)$ is taken.
 One then finds the appropriate diameter $d$ through
perturbation 
approach until 
$e^{\beta U_T(r)}g_T(r) = e^{\beta U(r)}g(r)$.
Where, subscript $T$ represents the trial system.
Once the hard sphere diameter $d$ is known the static
structure factor for
the LJ liquid is approximated by,
\begin{eqnarray}
S(q)=S_T(q)+\rho \int d {\bf r}[e^{\beta U_T(r)}
g_T(r)]
[e^{-\beta U(r)}-e^{-\beta U_T(r)}] e^{-i \bf q.r},
\end{eqnarray} 
For the trial hard sphere system the Percus-Yevick 
\cite{PY}
solution with Verlet-Wiess correction \cite{VW} is
used.

\section*{Power law exponents}
The ideal transition from an ergodic phase (fluid) to
a non-ergodic
(glass) phase is characterized in terms of
non-ergodicity parameter
defined as, $\psi_q(t\rightarrow\infty)=f(q)$.
In the long time limit or
$z \rightarrow 0$ limit of the Laplace-transformed
form of the
 equation (\ref{dynamic}) reduces to
\begin{equation}
\label{fullfq}
\frac{f(q)}{1-f(q)} = \int \frac{d^3k}{(2 \pi)^3}
[\tilde{V}^{(1)}(q,k)f(k)
+\tilde{V}^{(2)}(q,k)f(k)f(k_1)]
\end{equation}
\noindent
where vertices $\tilde{V}^{(1)}$ and $\tilde{V}^{(2)}$ are given by
equation (\ref{v1}) and 
equation (\ref{v2}). 
These set of coupled equations are solved
self-consistently for $f(q)$.
 In solving the integral equations a maximum cutoff
for the
wave vector integration $\Lambda \sigma = 30.0$
 with 200 grid points is used.
 A set of critical parameters
 ($\kappa_c, y_c, \rho_c, T_c$) are identified
below which all $f(q)'s$ converge to zero  
values (ergodic phase) and above that critical point
all $f(q)'s$ converge 
to nonzero values (nonergodic phase).
The critical points ($y_c$, $\kappa_c$, $\rho_c$,
$T_c$) in 
$T-\kappa$ phase space are located by fixing
$y_c=0.15$ and
varying $\kappa$ and  $T$.
Here we study the equations with the thermodynamic
parameters
changing in a way so that the volume change with the
temperature keeping
the Pressure constant, $P^* = P \sigma^3 /\epsilon =
0.5$ as shown in Fig. 1.
Then, {\em at a fixed pressure}, we observe
 a line of transition giving different critical temperatures
with different values of the metastability parameters.
In the ideal model under similar conditions
 a single transition point is obtained.
Along this transition line at a fixed pressure we
calculate, 
from the solutions of the MCT equations, the power law
exponent for 
relaxation in the LJ fluid as a function of
temperature and density.
We study the relaxation over time scales longer than
the microscopic 
time scales $t>(\Omega_q \tau)^{-1} $, where
$\tau$ is the 
 Lennard-Jones time
${(m\sigma^2/\epsilon)}^{\frac{1}{2}} $,
 but shorter compared to the $\alpha$-relaxation time
 scales.  
The dynamical equation (\ref{dynamic}),
for density correlation function $\psi_q(t)$ is solved
self consistently along the transition line shown in
Fig. 1.  
The  metastable potential well for defects used for
computing vertex 
functions  is shown in Fig. 2
at critical parameters ($T_c=0.544, y_c=0.15,
\kappa_c=-0.038$).
The observed relaxation behavior is given by,
\begin{equation}
\label{psi-dy}
\psi_q(t)=f_c(q)+C h(q)(t/t_o)^{-a}
\end{equation}
\noindent
where C is a constant and $t_o$ is the time scale over
 which the power law relaxation persists. 
In Fig. 3 we show the decay of density correlation
function with time 
($t/\tau)$.
The fit to the power law exponent is shown by dotted
lines and
the solid line represents the result from MCT
equations for temperature 
$T_c=.544$ in the power law regime.
Temperature dependence of the power law exponent $a$
is  
studied at a constant pressure $P^*=0.5$. 
The exponent $a$ is calculated at the transition
temperatures
for fixed $y_c=0.15$ and the corresponding
$\kappa_c's$.
Fig. 4 shows the variation of the power law exponent
$a$ with temperature.
We also calculate the wave vector dependence of
the power law exponent $a$. 
Fig. 5 shows the exponent $a$ along the wave vector
$q\sigma$ at $T=0.544$ and $\rho=0.958$.
In the inset the static structure factor is shown to
depict the alignment
of the maxima of $S(q\sigma)$ with minima of exponent $a$ and
vice-versa.
The arrows are shown at the maxima and  minima.
 The power law exponent is minimum at the
peak of the static structure factor, which reflects
the characteristic 
 slow relaxation at the diffraction maximum for
liquids.

\section*{Discussion}
In the simple mode coupling theory 
there is a single transition point with a fixed $T_c$.
As a result of
which the power law exponent determined in its
vicinity is temperature
 independent.
In the present case through the coupling of the
density
 fluctuations with defect motions a line of dynamic
instability
 is predicted and this results in a temperature
dependent
 exponent.
 Here we consider the simple one component LJ
 system and we obtain an oscillatory behavior of the
power law
 exponent a - as shown in Fig. 4.
For hard sphere system a similar oscillatory behavior
is found \cite{pre} from the solutions of the MCT
equations.
This was also shown
 to agree with the power law exponent that was
calculated from fitting 
of the light scattering data on
Colloids\cite{van-pusey}.
 A similar behavior was observed in the wave vector
dependence of
 the exponent as well.  
Computer simulations of one component systems at
metastable
 densities are particualrly 
 rare due to problem of crystallization and at this
time we are
 unable to find simulation data to look for such
behavior. 
 However simulation results for binary mixtures exist
and we studied
 the data reported in ref. \cite{kob} without
restricting
 to a fixed non-ergodicity parameter (NEP) as in the
ideal model
 that gives rise to a single exponent. We do observe in this case a 
 qualitatively same oscillatory behavior for the exponent $a$ of the 
 simulation data as shown in Figure 6.
 The NEP that is obtained in our fitting also shows qualitatively 
 similar behavior as predicted from the extended model studied here.
The present model can be extended to the $\alpha$-relaxation regime as well 
where the combined role of the density fluctuation and the defect 
fluctuation need to be considered.

\vspace*{1cm}
\section*{Acknowledgement}
SS acknowledges the support of University Grants
Commission (UGC)
 of India for the research.

\vspace*{2cm}

\section*{Figure Captions}

\vspace*{.5cm}
\noindent
Fig 1 :
Phase diagram in $T-\kappa$ space at $y_c=0.15$ along
a constant
pressure line. Temperature is expressed in units of
$\epsilon/k_B$.

\vspace*{.5cm}
\noindent
Fig 2 :
Variation of depth of the potential with $n/n^*$ for
$T_c^*=0.544$.
In the figure $h^*(n)$ represent the dimensionless
quantity 
$h(n)\beta \epsilon n \sigma^3$.

\vspace*{.5cm}
\noindent
Fig 3 :
Decay of normalized density-density correlation function $\psi_q(t/\tau)$ with
time ($t/\tau$)
 at $T=0.544$. Where $\tau$ is the 
 Lennard-Jones time
${(m\sigma^2/\epsilon)}^{\frac{1}{2}} $. Dotted lines 
show the power law fit (\ref{psi-dy}) (see text).

\vspace*{.5cm}
\noindent
Fig 4
Variation of power law exponent $a$ as a function of
temperature $T$.
$T$ is expressed in units of $\epsilon/k_B$.

\vspace*{.5cm}
\noindent
Fig 5 :
Variation of exponent $a$ as a function of wave number
$q\sigma$ for a packing fraction $T=0.544$. In inset
static structure 
factor is shown along wave number $q\sigma$. 

\vspace*{.5cm}
\noindent
Fig 6 : 
Variation of power law exponent $a$ with  temperature
as obtained by
fitting \ref{psi-dy}(see text) to the self correlation
function data from 
Ref. \cite{kob} at wave vector $q=7.22$.


\newpage
\vspace*{1cm}
\begin{thebibliography}{99}
%
\bibitem{freevol} M. H. Cohen and G. S. Grest, Phys.
Rev. B
{\bf 20}, 1077 (1979).

\bibitem{forster} D. Forster, {\it Hydrodynamic
Fluctuations,
Broken Symmetry and Correlation Functions}, (Benjamin,
Reading,
Mass., 1975).

\bibitem{kimmaj}
B. Kim and G. F. Mazenko, Phys. Rev. A {\bf 45}, 2393
(1992).

\bibitem{sid-trans}{\it Transport Theory and
Statistical Physics }
edited by S. Yip, (1995).

\bibitem{leshouc} W. G\H{o}tze in {\it Liquids,
Freezing and
the Glass Transition,} edited by D. Levisque, J. P.
Hansen,
and J. Zinn-Justin (Elsevier, New York, 1991);


\bibitem{ullo-yip}
J. Ullo and S. Yip, Phys. Rev. Lett. {\bf 54}, 1509
(1985);
Phys. Rev. A. {\bf 39}, 5877 (1989).

\bibitem{hiwa}
Y. Hiwatari, J. Phys. C {\bf 13}, 5899 (1980).


\bibitem{megen} W. van Megen, T. C. Mortensen, S. R.
Williams and J. 
M\"{u}ller, Phys. Rev. E {\bf 58}, 6073 (1998).

\bibitem{van-uwood} W. van Megen, and S. M. Underwood
Phys. Rev. E
 {\bf49}, 4206 (1994). 

\bibitem{van-pusey} W. van Megen, and P. N. Pusey
Phys. Rev. A
 {\bf43}, 5429 (1991). 

\bibitem{beng} U. Bengtzelius, W. Gotze and A.
Sj\"{o}lander
Phys. Rev. E {\bf 50}, 1265 (1994).


\bibitem{leu}
E. Leutheusser Phys. Rev. A {\bf 29}, 2765 (1984)

\bibitem{D+M}
S.P.~Das, G.F.~Mazenko, 
Phys.~Rev.~A.~{\bf 34}, 2265 (1986).

\bibitem{jim}
R. Schimitz, J. W. Dufty and P. De,
Phys.~Rev. Lett. {\bf 71} 2066 (1993).

\bibitem{gt-sj}W. Gotze and L. Sj\"{o}gren
Rep. Prog. Phys. {\bf 55}, 241 (1992).

\bibitem{dsch}S. P. Das and R. Schilling
Phys. Rev. E {\bf 50}, 1265 (1994).

\bibitem{spd}
S. P. Das and S. Srivastava,
Phys. Letters A, {\bf 266} 58 (2000).

\bibitem{yeomaj}J. Yeo and G. F. Mazenko,
 J. of Non-Crystalline Solids, {\bf 172-174}, 1(1994),
Phys. Rev. E {\bf 51}, 5752 (1995).

\bibitem{yeo} J. Yeo, Phys. Rev. E {\bf 52}, 853
(1995).

\bibitem{beng1} U. Bengtzelius, 
Phys. Rev. A {\bf 34}, 5059 (1986).
\bibitem{kirk}
T. R. Kirkpatrick, Phys. rev. A. {\bf 32}, 3120
(1985).

\bibitem{abramowitz} M. Abramowitz and I. Stegun, {\it Handbook of
Mathematical Functions}
(Dover, New York, 1965);

\bibitem{PY}
J. P. Hansen and J.R. McDonald, {\it Theory of Simple
Liquids}
 (Academic, London, 1976).

\bibitem{bongsoo} B. Kim Phys. Rev. A {\bf 46}, 1992
(1992).

\bibitem{wca}
J. D. Weeks, D. Chandler and H. C. Andersen,
J. Chem. Phys. {\bf 55}, 5422 (1971).

\bibitem{anderson} 
H. C. Anderson, D. Chandler and J. D. Weeks, 
J. Chem. Phys. {\bf 56} 3812 (1972).

\bibitem{VW}
L. Verlet and J. J. Weiss, Phys. rev. A. {\bf 45}, 898
(1992).

\bibitem{pre}
S. Srivastava and S. P. Das, to be published in
Physical Review E.

\bibitem{kob}
T. Gleim and W. Kob, Eur. Phys. J. B. {\bf 13}, 83
(2000).

\end{thebibliography}
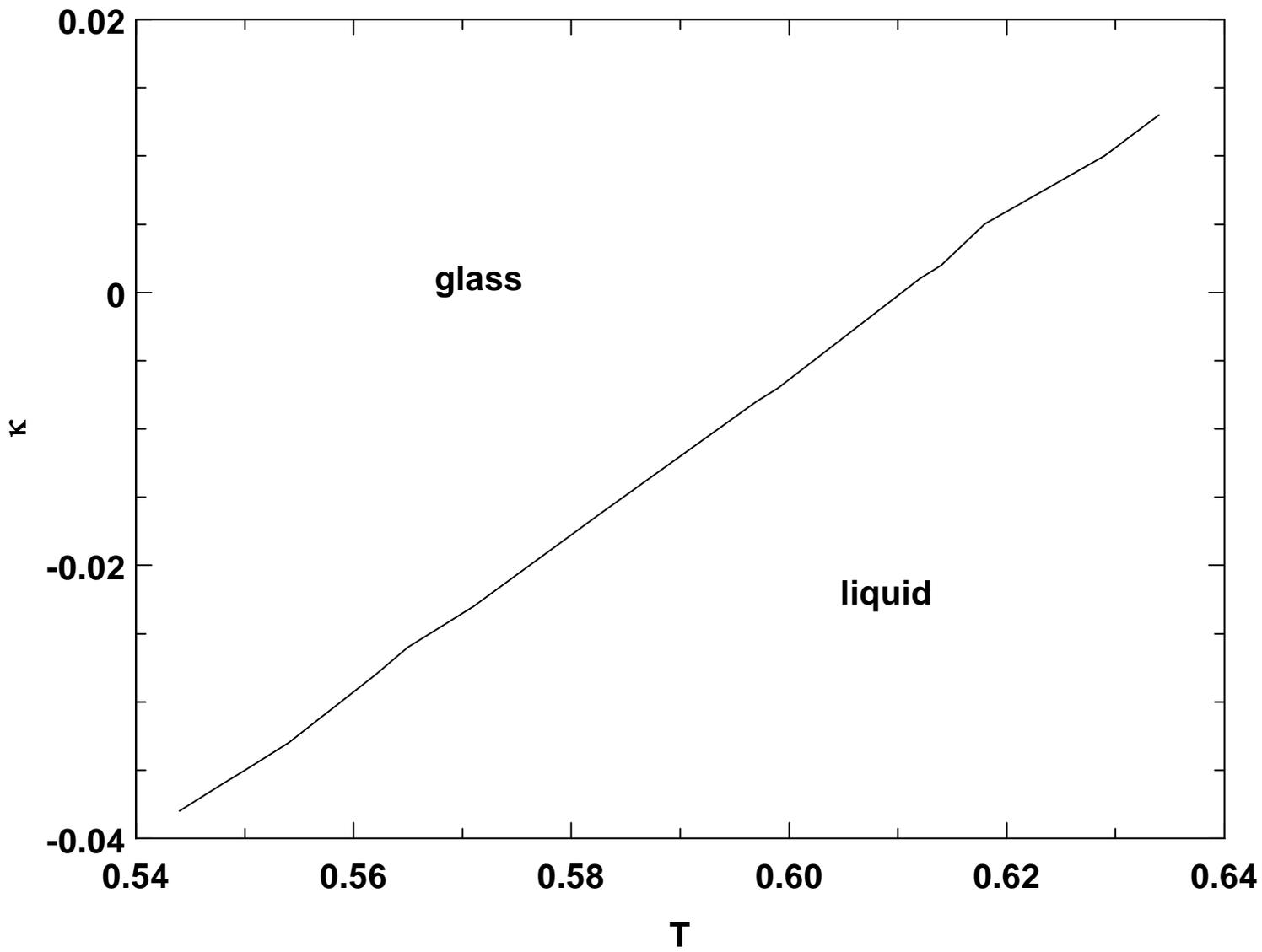
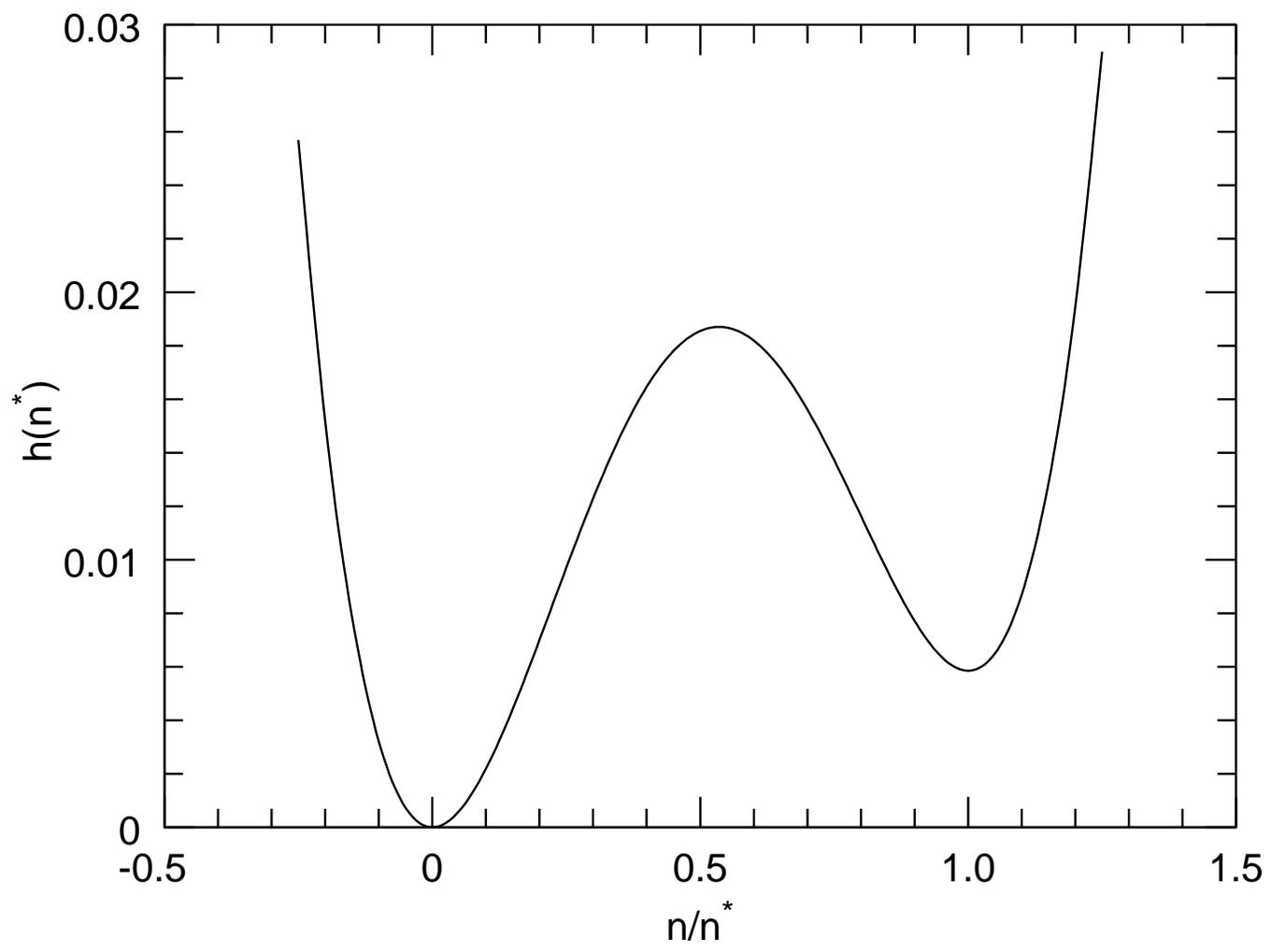
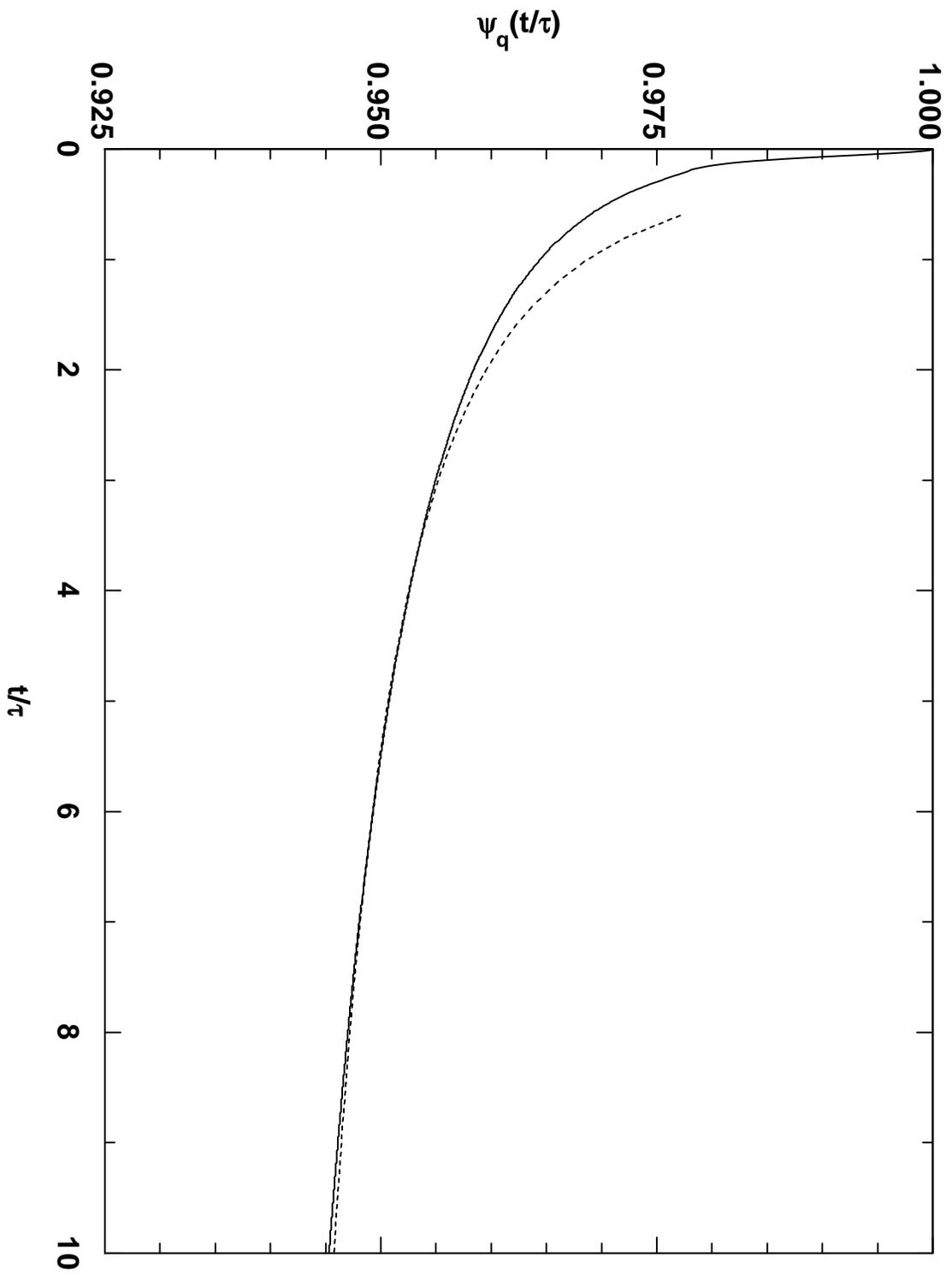
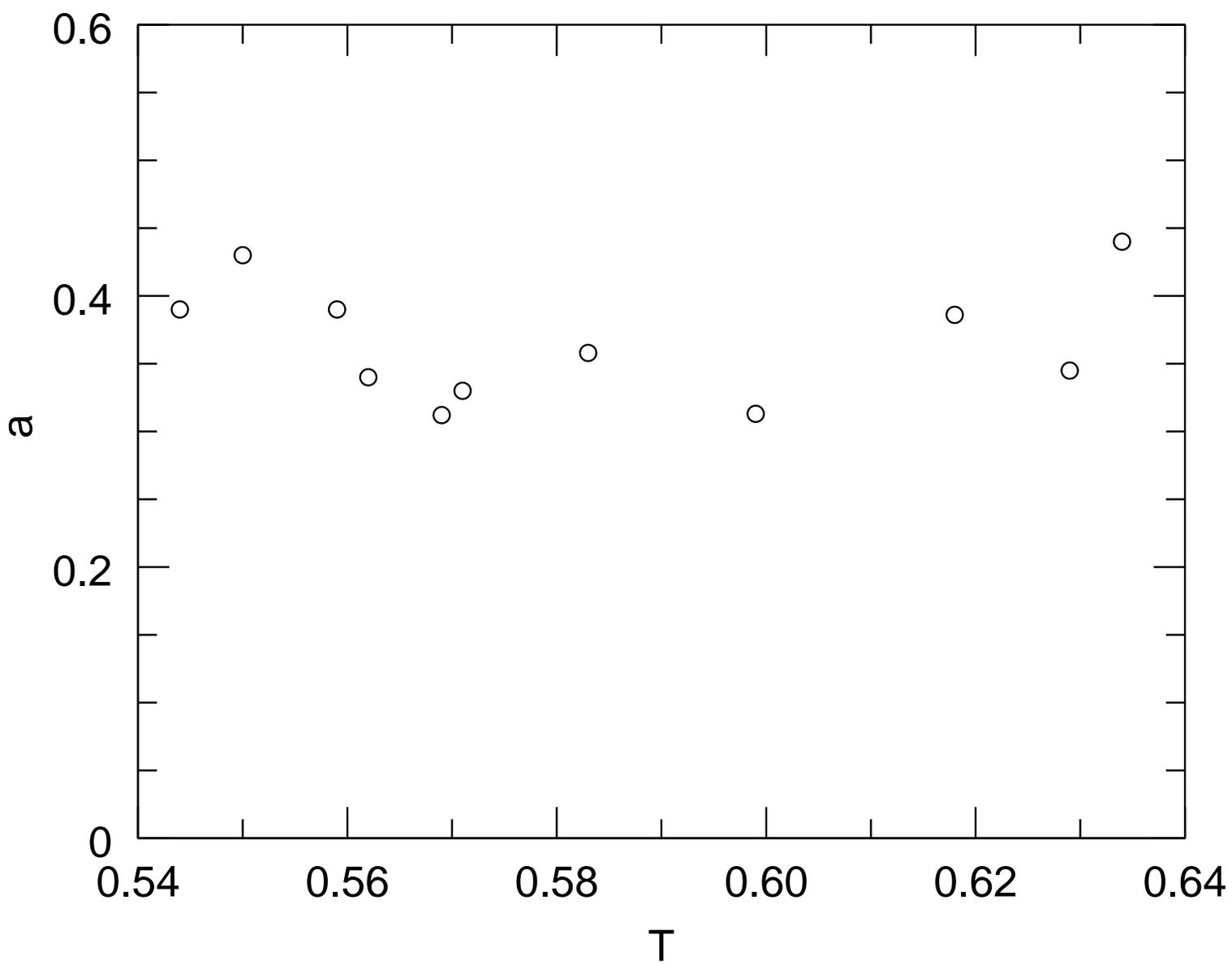
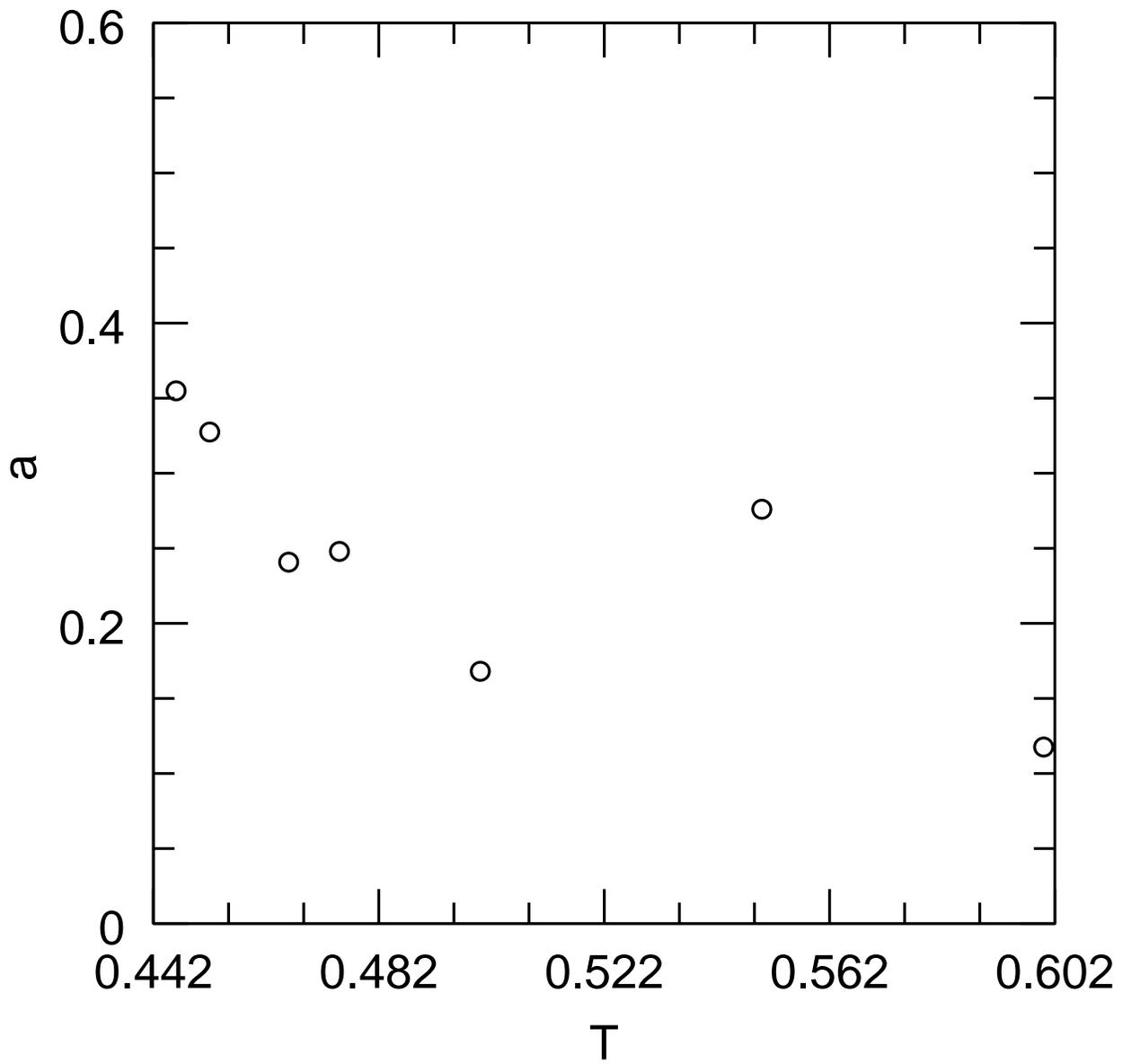
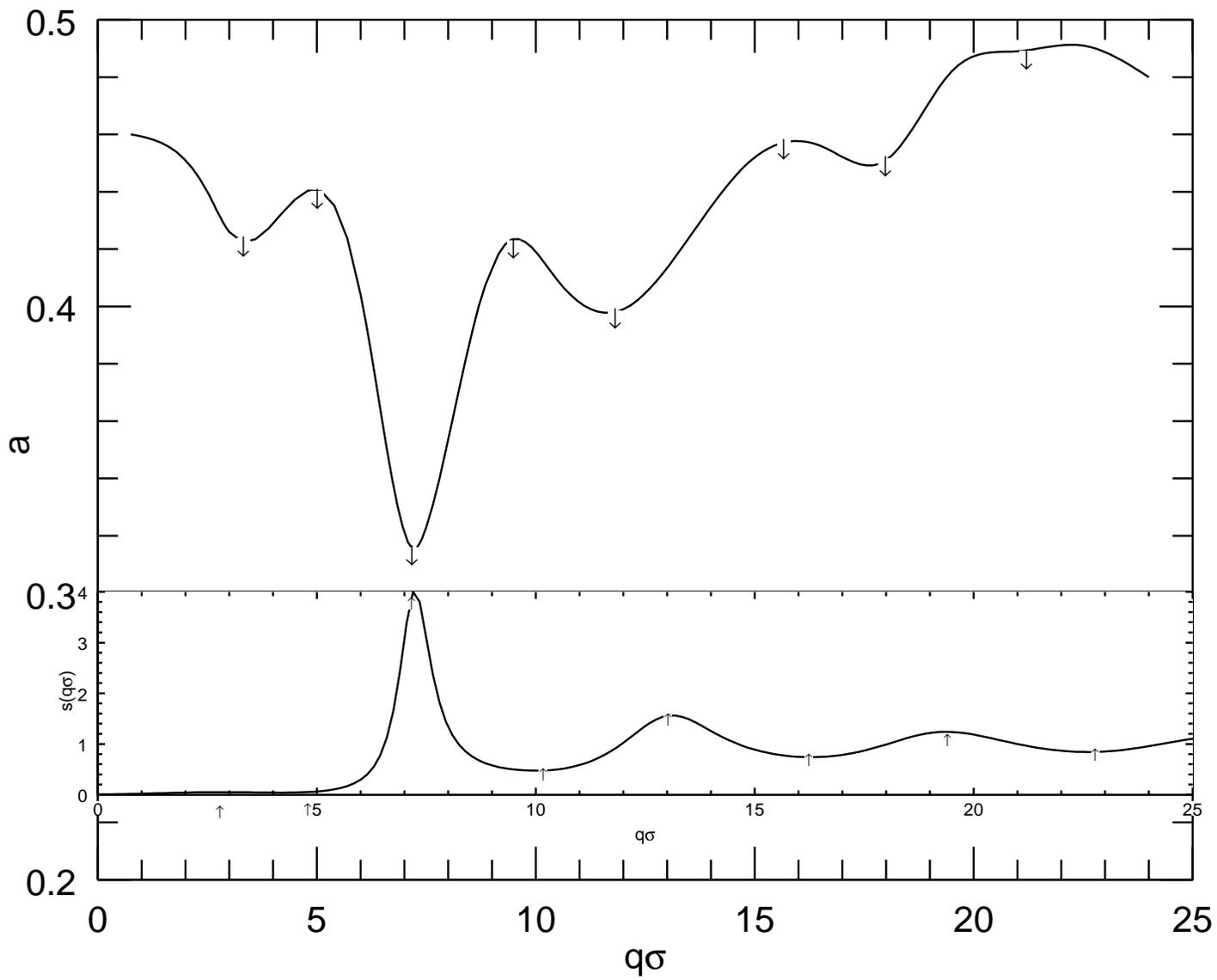
\end{document}